\documentclass[conference]{IEEEtran}
\pdfoutput=1
\usepackage{graphicx}
\usepackage{hyperref}
\hypersetup{
    colorlinks=true,
    linkcolor=black,
    citecolor=black,
    filecolor=black,
    urlcolor=black,
}

\begin{document}

\title{Using Calculation Fragments for \\Spreadsheet Testing and Debugging}

\author{\IEEEauthorblockN{Dietmar Jannach,
Thomas Schmitz}
\IEEEauthorblockA{Department of Computer Science, TU Dortmund, Germany\\
dietmar.jannach@tu-dortmund.de, thomas.schmitz@tu-dortmund.de}}

\maketitle

\begin{abstract}
A number of automated techniques and tools were proposed in the research literature over the years which aim to support the spreadsheet developer in the process of testing and debugging a faulty spreadsheet. One underlying assumption of many of these approaches is that the spreadsheet developer is capable of providing test cases or is at least reliably able to determine whether a calculated value in a certain cell is correct given the current set of inputs.

Since real-world spreadsheets can be complex, we argue that these assumptions might be too strong in some situations. We therefore propose to support the user during testing and debugging by automatically computing \emph{spreadsheet fragments} of manageable size. The spreadsheet developer can then verify the correctness of a smaller set of formulas for which the calculated output can be more easily validated.

\end{abstract}

\section{Introduction}
In environments like MS Excel, the set of built-in features designed to support the users when testing or debugging their spreadsheet programs is quite limited. Examples of the few available features include the visualization of individual cell dependencies or the markup of cells containing suspicious formulas (``smells'' in the sense of \cite{abreu2014smelling} or \cite{Hermans2012a}).  Over the last decades a variety of different tools and techniques were proposed in the research literature to help the user avoid, locate and remove errors in spreadsheets~\cite{Jannach2014JSS}. The proposed approaches range from intelligent visualizations over test case generation to the application of novel debugging techniques.

Many of these testing and debugging approaches assume that the users are able to provide test cases, which contain the expected values for the output cells of their spreadsheets \cite{Abraham:2007:GSD:1248820.1248858, JannachEtAl2014ASE}; some techniques require the user at least to reliably indicate if an output cell value or a test case is correct or faulty \cite{Abraham:2006:ATA:1174509.1174656,Hofer2013EEF}.
Providing expected values or even assessing the correctness of individual values might, however, be challenging for the user in particular when the spreadsheets are large and when no known-to-be-correct test cases are available.

We therefore propose to (automatically) modularize the spreadsheet under investigation and ask the user for feedback on the correctness of calculations of smaller and more comprehensible spreadsheet \emph{fragments}. These fragments in some sense correspond to the concept of \emph{unit tests} in standard software development processes based on which the correctness of smaller functional parts of the program can be validated.

The proposed fragment extraction approach is also related to automated refactoring techniques for long methods in imperative programs based, e.g., on control flow and data flow graphs like in \cite{Maruyama:2001:AMR:375212.375233}. However, spreadsheets have no standard control flow graphs and the data flow graphs are much simpler, because the formulas in the cells do not directly change the values of other cells.
Existing refactoring approaches therefore cannot be directly applied and further investigations are required to assess how these methods can be adapted to be applicable for the special and defined structure of spreadsheets.

\section{Example}
\begin{figure}[tbh]
	\centering
	\includegraphics[width=0.477\textwidth]{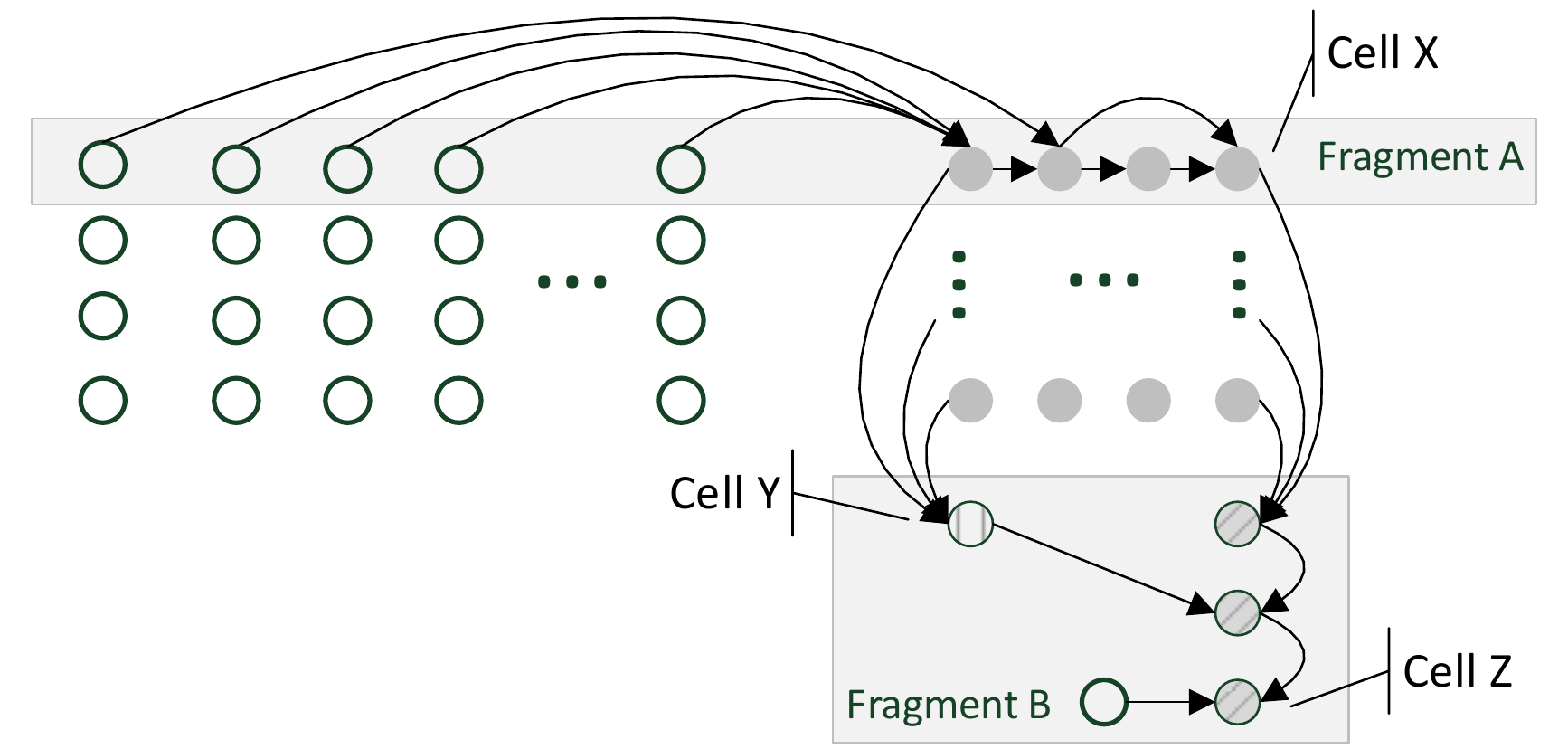}
	\caption{A typical spreadsheet with fragments.}
	\label{fig:figure_1}
\end{figure}

Consider the example in Figure \ref{fig:figure_1}, which shows the dependency structure of a typical financial calculation sheet adapted from \cite{JannachEtAl2014ASE}. In the upper part of the spreadsheet (Fragment A), monthly sales data are aggregated with the help of a number of copy-equivalent rows. In the lower part (Fragment B), further data aggregation is done and additional calculations are made on the aggregate values.
Structurally or semantically different cells are denoted by differently styled circles, e.g., the empty circles denote input cells or fixed parameters.

Let us assume that the spreadsheet developer observes that the ``final outcome'' in the lower right part of the spreadsheet (Cell Z) only meets his expectations for some but not all tested input value constellations. However, providing the exact expected values -- as required by some automated debugging methods \cite{Abraham:2007:GSD:1248820.1248858, JannachEtAl2014ASE} -- might be cumbersome for the user and error-prone in particular when the spreadsheet is complex.

In this example, we could -- as indicated in Figure \ref{fig:figure_1} -- have two fragments that can be tested individually and in the case of Fragment A use one row of the spreadsheet as a representative for the other copy-equivalent rows.
These fragments could be defined manually by the user; in our work, we are, however, interested in techniques to automatically identify possible fragments and provide adequate tool support.

\section{Design Considerations - Technical Approach}

\subsection{End User Perspective}
As spreadsheet developers are usually not IT experts or programmers, special care has to be taken when designing a tool for fragment-based testing and debugging. We therefore plan to integrate the sketched techniques in our model-based debugging plug-in to Excel called \textsc{\small{Exquisite}} \cite{JannachEtAl2014ASE}. When using this tool, the users can stay within their usual spreadsheet environment; appropriate user-oriented metaphors for those concepts that are not part of typical spreadsheet tools (like test cases, fragments, or unit tests), however, still have to be found and evaluated with users.

Another open question in that context is related to the optimal complexity of the fragments a user should work on. If the fragments are too small, test cases for too many fragments have to be defined by the user; if they are complex and span major parts of the spreadsheet, the cognitive effort for the user when providing the test cases might be too high.

\subsection{Automated Extraction of Fragments}
The main goal of our work is to develop algorithms and heuristics to automatically extract potentially overlapping fragments that are manageable in size and in the best case semantically connected. These fragments can then for example serve as a basis for the manual or automated creation of test cases, e.g., using property-based testing techniques \cite{Fink:1997:PTN:263244.263267}, where the goal is to automatically create test cases which falsify user-specified properties of the output values. In another scenario, the fragments can be used to inspect parts of the spreadsheet which are considered to be ``suspicious'' as a result of a diagnosis or smell detection technique.
We are currently exploring the following fragment construction strategies.

\subsubsection{Collapsing Copy-Equivalent Structures}
In Figure \ref{fig:figure_1}, only the last four cells on the right of Fragment A (including Cell X) contain formulas. When the right-most cells of each row are considered suspicious, it can be sufficient to create a fragment which comprises \emph{one representative row}, so that the user only has to create one test case for all rows.

On the other hand, the left-most cell of Fragment B (Cell Y) is based on the values of several copy-equivalent cells and, e.g., corresponds to the sum of the monthly sales figures. If we defined this cell and a subset of its inputs as another Fragment C (not shown in Figure \ref{fig:figure_1}), the fragment could contain two or more of the copy-equivalent rows such that the user can validate the correctness of the aggregation function, which would be impossible when there is only one input row. With this second technique the user could, however, miss range errors such as an omitted cell of a sum.

\subsubsection{Limiting the Dependency Paths}
Fragments of manageable size like Fragment B in Figure \ref{fig:figure_1} can be obtained by limiting the depth and maybe also the breadth of the dependency paths leading to a suspicious cell. The depth limitation could be based on simple path length restriction heuristics or based on structural or semantic considerations, e.g., by observing that we encounter a set of copy-equivalent cells. Existing techniques developed for spreadsheet visualization and comprehension could be applicable in that context.

\subsection{Interactive Testing and Debugging}
The provision of adequate tool support when testing or debugging a calculation fragment is finally a crucial part of our approach. From a UI perspective, one could for example create a new worksheet which only contains the cells and formulas of the fragment under investigation. This could however remove context information, as neighboring cells outside the fragment would not be visible. Therefore, we propose to only visually dim the cells outside the fragment and make them read-only to help the user focus on the current task.

For debugging purposes, our model-based diagnosis approach presented in \cite{JannachEtAl2014ASE} can be used to find the possible causes whenever there is a discrepancy between the expected and the calculated value of an output cell of a fragment. Because of the small sizes of the fragments, we conjecture that the combination of these approaches could help to quickly find the faulty formulas.

In addition, the proposed test and debugging environment should be able to automatically create appropriate test cases which are easy to validate manually. Again, existing approaches for test case generation from the literature should be applied. Note that when creating these test cases, input values have to be generated only for the cells at the fragment borders (e.g., Cell Y in Fragment B). The developed software environment and Excel plug-in has furthermore to be extended in a way that test cases can be easily stored, adapted and automatically executed, e.g., for regression testing.

\section{Summary}
We argue that providing exact values for expected calculation outcomes in spreadsheet testing and debugging scenarios might be too challenging for users in some situations. We therefore propose to further investigate approaches in which the given spreadsheet is decomposed into smaller fragments which can be more easily validated by the user.

\bibliographystyle{abbrv}

\end{document}